\documentclass[useAMS,usenatbib]{mn2e}
\usepackage{graphicx}
\usepackage{amssymb}

\newcommand{\etal}{et al.~}
\def\hst{{\sl {\it HST}}}
\def\rxte{{\sl RXTE}}

\def\XMM{{\sl XMM-Newton}}
\def\xmm{{\sl XMM-Newton}}

\def\chandra{{\sl Chandra}}

\def\gtrsim{\lower 2pt \hbox{$\, \buildrel {\scriptstyle >}\over
{\scriptstyle \sim}\,$}}
\def\lesssim{\lower 2pt \hbox{$\, \buildrel {\scriptstyle <}\over
{\scriptstyle \sim}\,$}}

\def\heii{He~{\scriptsize II}}

\def\oviii{O~{\scriptsize VIII}}
\def\ovii{O~{\scriptsize VII}}
\def\ovi{O~{\scriptsize VI}}

\def\neix{Ne~{\scriptsize IX}}

\def\nv{N~{\scriptsize V}}

\def\ka{K{$\alpha$}}

\usepackage{fixltx2e}
\begin{document}

\title[Finding the Missing OVII K$\alpha$ Absorption Line]{{\it{XMM-Newton}}/RGS Detection of the Missing Interstellar OVII K$\alpha$ Absorption Line in the Spectrum of Cyg X-2}
\author[]{Samuel H. C. Cabot$^{1}$, Q. Daniel Wang$^{2}$\thanks{E-mail:wqd@astro.umass.edu}, and Yangsen Yao$^{3,4}$\\
$^{1}$Berkshire School, Sheffield, MA 01257, USA\\
$^{2}$Department of Astronomy, University of Massachusetts, 
  Amherst, MA 01003, USA\\
$^{3}$Center for Astrophysics and Space Astronomy, Department of Astrophysical and Planetary Sciences, University of Colorado, Boulder, CO 80309\\
$^{4}$Eureka Scientific, Oakland, CA 94602, USA}


\maketitle

\begin{abstract}
The hot interstellar medium is an important part of the Galactic ecosystem and 
can be effectively characterized through X-ray 
absorption line spectroscopy. However, in a study of the hot medium using 
the accreting neutron star X-ray binary, Cyg X-2, as a background 
light source, a mystery came about when the 
putatively strong \ovii\ K$\alpha$ line was not detected in {\it{Chandra}} grating observations, while other normally weaker lines such as \ovii\ K$\beta$ 
as well as \ovi\ and \oviii\ K$\alpha$ are clearly present (Yao et al. 2009). 
We have investigated the grating spectra of Cyg X-2 from 10 {\it{XMM-Newton}} 
observations, in search of the missing line. We detect 
it consistently in nine of these observations, but
the line is absent in the remaining one observation or is inconsistent with the 
detection in others at a $\sim 4\sigma$ confidence level. This absence of the line
resembles that seen in the {\it Chandra} observations. 
Similarly, the \ovi\  
K$\alpha$ line is found to disappear occasionally, but not in concert with the
variation of the \ovii\ K$\alpha$ line. 
All these variations are most likely due to the presence of changing 
\ovii\  and \ovi\ K$\alpha$ emission lines of Cyg X-2, which
are blurred together with the absorption ones in the X-ray spectra. 
A re-examination of the {\it Chandra} grating data indeed shows
evidence for a narrow emission line slightly off the \ovi\ \ka\
absorption line. We further show that narrow \nv\ emission
lines with varying centroids and fluxes are present in far-UV spectra from the Cosmic Origins Spectrograph 
aboard the Hubble Space Telescope. 
These results provide new constraints on the accretion 
around the neutron star and on the X-ray-heating of the stellar companion.
The understanding of these physical processes is also important to the fidelity
of using such local X-ray binaries for interstellar 
absorption line spectroscopy.

\end{abstract}


\section{INTRODUCTION}

Diffuse hot gas is an important, yet probably the least understood component of the interstellar medium (ISM) 
in and around the Milky Way (e.g., Wang 2010; Putman \etal\ 2012 
and references therein). The hot ISM has been proposed to account for much of the missing baryon matter within the 
Galactic halo and traces energetic feedback from stars, which shapes the
ecosystem of the Galaxy. However, little is actually known observationally 
about the kinematic, chemical and spatial properties, 
as well as the content of the hot ISM. 
It has been chiefly studied in its emission, which carries 
little distance information and is subject to absorption by 
cool gas -- an effect that is often difficult to quantify.
In the past several years, however, it has been demonstrated that the hot ISM can be effectively characterized 
through X-ray absorption line spectroscopy (e.g., Wang \etal\ 2005; Williams \etal\ 2005; Yao
\& Wang 2005; 	Juett et al. 2006; Costantini \etal\  2012; Wang 2010 and references therein). Through the measurements of absorption lines 
produced by various ions in the spectra of bright X-ray sources, primarily active galactic nuclei and X-ray 
binaries, it becomes possible to constrain the spatial, physical, chemical, and kinematic properties of the hot gas. 
One key step in such a study is the identification and correction for any potential spurious features, 
which may be intrinsic to the background source or to the instruments. This issue is particularly acute for local 
sources, which may produce features that cannot be easily recognized because of little systematic velocity offset. 

The existing interstellar X-ray absorption line study of Cyg X-2 represents such an outstanding case 
(Yao et al. 2009). 
This archetypal Z-source is a bright binary system containing a neutron star, accreting from 
its low-mass companion (V1341) with an orbital period of 9.84450 days (Casares et al. 2009).
In their study with a {\it{Chandra}} grating spectrum, Yao et al. (2009) 
particularly noted a puzzling lack of the interstellar \ovii\ K$\alpha$ (21.602~\AA) absorption line. 
Other important Oxygen absorption lines, such as \ovii\ K$\beta$ (18.654~\AA), 
\oviii\ K$\alpha$ (18.967~\AA) and \ovi\ K$\alpha$ 
(22.040~\AA), were all detected. The missing line is known to lay in the middle of these other lines 
in terms of the transition wavelengths and ionization states of the ions and is usually known to be the
strongest of them all. One possibility presented, but ruled out, by Yao et al. was that 
the absorption line might be filled or contaminated by a broad emission line similar to those detected at wavelengths shorter than 10~\AA\ 
in the same spectrum (Schulz et al. 2009). These broad emission lines originate from 
the photo-ionized accretion disk corona (ADC) surrounding the neutron star. 
The presence of a broad \ovii\ K$\alpha$ emission line, which could only be partly absorbed, would leave 
broad wings in the spectrum, which were not detected. 

In the present work, we test a hypothesis that the absorption line could be contaminated by a 
relatively narrow emission line. 
Such a line or lines might be expected from X-ray heating of the atmosphere of 
the companion star and/or an outer region of the expected accretion disk, where the Doppler broadening may be sufficiently small. 
This is possible because the \ovii\ K$\alpha$ line is 
at a lower energy than where the broad emission lines were detected.
At such lower energies, 
relatively narrow emission lines were indeed observed 
in the spectra of LMC X-3 (Song et al. 2010), for example, although it is a high-mass X-ray binary with an accreting black hole. If this was the case, 
Cyg X-2 may then be a problematic background source for the study of interstellar absorption lines. 
A varying, narrow emission 
line could not only make absorption lines 'vanish', but change their depths and other properties key to such a study. In the present work,
we mainly use multiple {\it{XMM-Newton}} observations to test our 
hypothesis and to potentially 
enhance our understanding of the hot ISM along the line of sight as well as this well-known source itself.

Indeed, Vrtilek et al. (2003) have suggested that \heii\ and \nv\ emission lines, as detected with {\sl HST}/GHRS observations of Cyg X-2, seem to arise from the surface of the companion. These lines show velocity shifts that are consistent with this interpretation; but the FWHMs of the lines are larger than expected. It may be possible that the line emission is from an accretion heated corona above the disk; however, the FWHMs are smaller than expected by simple models of line emission from accretion disks; the lines also do not exhibit the double-peaked structure as predicted. We use recently available {\sl {\it HST}}/COS
data of Cyg X-2 with improved sensitivity and resolution to search for the \nv\ counterpart of the
putative \ovii\ \ka\ line and to determine the nature of the line emission.
In addition, we have carefully re-examined the \chandra/HETG data to
check the consistency of the \XMM\ results and to further the test of
our scenario. 

The organization of the paper is as follows: In \S~2, we describe the {\it{XMM-Newton}} observations
and our data reduction procedure. We present the analysis and results of the data in \S~3. The results are
then compared with those from other X-ray and UV data analyses in \S~4. In \S~5, we 
discuss the origins of the absorption and emission seen in the Cyg
X-2, as well as the consistency of our scenario. All errors are measured 
at the 68\% (1$\sigma$) confidence, unless marked otherwise.

\section{{\it{XMM-Newton}} OBSERVATIONS AND DATA REDUCTION}

\begin{table*}
\tiny
  \caption{{\sl XMM-Newton} Observations.}
  \begin{tabular}{@{}lcrrccccccccr@{}}
\hline\hline
Obs\#&ObsID &  $t_o$ (s) & $t_c$ (s) & Start Time & $\phi$ & $f_{ASM}$
& CR & $\chi^2/\nu$ & $\lambda_{ovii}$ (\AA) & $\lambda_{ovi}$ (\AA) &  EW$_{ovii}$ (m\AA) & EW$_{ovi}$ (m\AA)\\
\hline  
N1& 0111360101 & 21481 & 14030 & 2002-06-03@10:13 & 0.83 & 33.84 & 1.03 & 82.36/79 & ${21.6265}_{-0.0146}^{+0.0157}$ & ${22.0967}_{-0.0193}^{+0.0180}$ &  $20.6_{-5.0}^{+5.0}$ & $14.3_{-5.5}^{+5.5}$  \\
N2& 0303280101 & 31784 & 31510 & 2005-06-14@13:49 & 0.30 & 34.37 & 0.97 & 83.19/77 & ${21.6147}_{-0.0105}^{+0.0111}$ & ${22.0605}_{-0.0202}^{+0.0203}$ &  $20.1_{-3.4}^{+3.4}$ & $12.2_{-3.7}^{+3.7}$  \\
N3& 0561180201 & 10004 & 9753  & 2008-11-10@13:40 & 0.76 & 30.42 & 1.20 & 87.20/82 & ${21.6114}_{-0.0198}^{+0.0212}$ &  22.0254 (fixed)            &  $13.8_{-5.8}^{+5.8}$ & $2.9_{-2.9}^{+6.4}$   \\
N4& 0561180401 & 4942  & 4786  & 2008-12-22@05:40 & 0.99 & 35.67 & 1.11 & 77.27/81 &  21.6081 (fixed)            & ${22.0173}_{-0.0196}^{+0.0220}$ &  $0.0^{+4.9}$        & $21.9_{-8.4}^{+9.4}$  \\ 
N5& 0602310101 & 84850 & 83270 & 2009-05-12@09:44 & 0.37 & 34.56 & 1.00 & 110.8/76 & ${21.6132}_{-0.0098}^{+0.0079}$ & ${22.0377}_{-0.0163}^{+0.0135}$ &  $22.5_{-2.0}^{+2.0}$ & $9.9_{-2.3}^{+2.3}$   \\
N6& 0561180501 & 13725 & 13280 & 2009-05-13@10:03 & 0.44 & 30.14 & 1.02 & 96.97/81 & ${21.6081}_{-0.0157}^{+0.0142}$ &  22.0254 (fixed)           &  $19.5_{-5.2}^{+5.2}$ & $0.0^{+2.5}$   \\
N7& 0650040801 & 30941 & 3138  & 2010-11-22@13:24 & 0.22 & 46.18 & 1.37 & 81.96/78 & ${21.6239}_{-0.0351}^{+0.0288}$ & ${22.0402}_{-0.0173}^{+0.0171}$ &  $14.6_{-9.0}^{+8.9}$ & $29.9_{-9.1}^{+9.1}$  \\
N8& 0650040901 & 30940 & 15540 & 2010-11-24@05:47 & 0.39 & 35.51 & 1.30 & 102.8/79 & ${21.5965}_{-0.0099}^{+0.0194}$ & ${22.0090}_{-0.0337}^{+0.0261}$ &  $19.7_{-4.2}^{+4.1}$ & $8.5_{-4.6}^{+4.6}$   \\
N9& 0650041001 & 30939 & 3638  & 2010-11-24@15:00 & 0.43 & 37.16 & 1.24 & 81.71/78 & ${21.5830}_{-0.0128}^{+0.0128}$ & ${21.9814}_{-0.0200}^{+0.0236}$ &  $34.2_{-7.9}^{+7.8}$ & $16.3_{-10.9}^{+08.3}$\\
N10&0650041101 & 30939 & 30610 & 2010-11-26@13:15 & 0.62 & 17.33 & 1.16 & 79.87/77 & ${21.5885}_{-0.0154}^{+0.0128}$  & ${21.9899}_{-0.0202}^{+0.0207}$ &  $16.9_{-3.2}^{+3.1}$ & $9.1_{-3.4}^{+3.4}$   \\
N1-10 & Combined & 290545 & 209600 & - & - & - & 1.05 & 201.0/165                  & ${21.6081}_{-0.0035}^{+0.0036}$  & ${22.0255}_{-0.0089}^{+0.0094}$ &  $19.6_{-1.3}^{+1.3}$ & $8.6_{-1.4}^{+1.4}$   \\   
\hline
\label{t:xmm}
\end{tabular}

Our data analysis includes ten archival {\it{XMM-Newton}} exposures (Table~\ref{t:xmm}). 
Only the data from one of the two reflection grating spectrometers (RGS1) are used. 
The RGS2 had a failed CCD covering the wavelength range including the \ovii\ K$\alpha$ line, 
rendering the data irrelevant in the present study.

Note: $t_o$ and $t_c$ are the original and cleaned exposures, while $\phi$ is the Cyg X-2 orbital phase 
of an individual observation.
$f_{ASM}$ in units of $ASM$ $counts \rm~s^{-1}$ is the 1.5-12 keV band flux from RXTE/ASM
($http://xte.mit.edu/ASM_lc.html$). The RGS count rate (CR), in units of $counts \rm~s^{-1}$, is estimated in the range of 21.0-22.7~\AA. 
The $\chi^2$ and number of degrees of freedom are for the best model fit to the RGS spectrum.
The last row (N1-10) gives the parameters for the combined spectrum.
\end{table*}

We reduce the data, using the current calibration files and Scientific Analysis System software (version 10.0.0). 
Our spectral extraction from each observation adopts the source coordinates of 
Cyg X-2 (RA = $21^h44^m41.20^s$, Dec = 38$^\circ$19$^\prime$18.0$^{\prime\prime}$; J2000) 
from NED and the default 90\% count enclosed region. Both the background and 
response matrix files are also extracted 
using the default setting. We also produce a combined spectrum by merging the on-source
spectral, response matrix, and background files from individual observations. We fit the spectrum with 
a simple power law with two multiplicative Gaussian components representing the \ovii\ and \ovi\
K$\alpha$ absorption lines, as well as a foreground photo-electric absorption in the range 
of 21.0 - 22.7~\AA. Our focus is on the measurement of the absorption lines, 
which is insensitive to the 
specific value of the photo-electric absorption and the power law index
(the two are strongly correlated). 
Therefore, in subsequent
analysis we fix the absorption column density to $2.4 \times 10^{21} {\rm~cm^{-2}}$, which is the best fit to the 
combined spectrum. We also fix the intrinsic widths (the Gaussian dispersion $\sigma$) of the absorption lines
at $3.4 \times 10^{-4}$ keV as they are not resolved within the X-ray spectra.

\begin{figure}
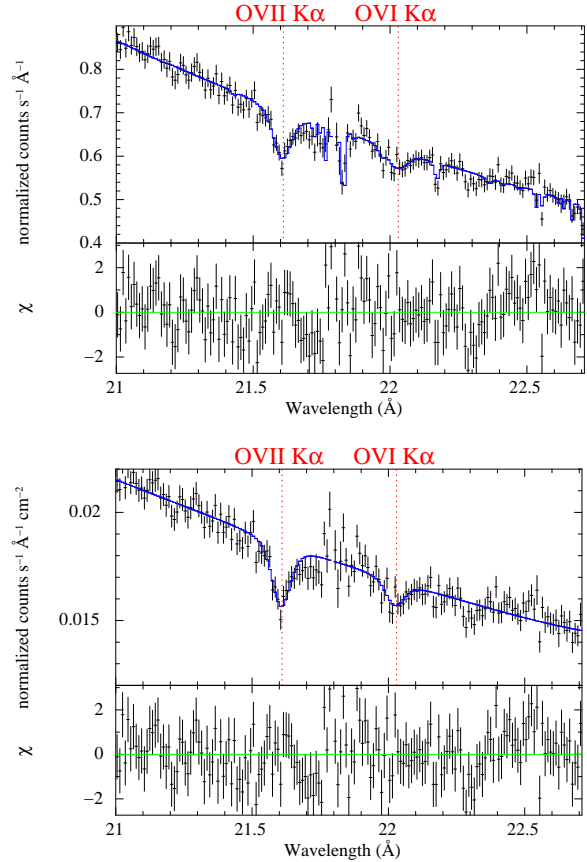

\includegraphics[width=2.2in,angle=270]{fig/f1a.ps}
\includegraphics[width=2.3in,angle=270]{fig/f1b.ps}
\caption{{\sl Upper panel:} Combined RGS1 spectrum of Cyg X-2 together with the 
best-fit (blue solid curve) model (\S~2). {\sl Lower panel:}
The same spectrum and fit, but approximately 
corrected for the effective area of the instrument.
The relative deviations from the fit are shown at the lower parts of each panel.}
\label{f:spec_c}
\end{figure}

A few spectral channels show substantially large deviations
from the best-fit model. These individual channels span wavelength intervals that are much smaller than the characteristic width of the line
response function (LRF); therefore, the deviations must be instrumental, which indeed tend to appear at 
edges of flagged bad columns or CCD gaps, where the effective area of the instrument
is not well calibrated. To systematically identify such channels, we compare the
spectrum from each observation 
with the same best-fit model, except for the power law normalization, which is refitted. We flag
channels with more than 3$\sigma$ deviations from the fit to be bad. This exclusion of the channels 
provides an acceptable balance between removing affected data and keeping good ones. 
The resultant spectra are merged again to form the final combined spectrum, which is then 
fitted with the same model. This fit is presented in Fig.~\ref{f:spec_c}, while the interested model 
parameters are included in Table~\ref{t:xmm}. 

A slightly different processing procedure is also tried, following the referee's 
suggestion. While our data processing uses a default keepcool=yes
setting of the SAS tool {\sl rgsproc}, this setting does not let {\sl rgsproc} 
detect bad pixels. To test whether or not a different setting would make a 
significant difference, we perform a comparison of 
centroids and equivalent widths (EWs) for both the \ovii\  and \ovi\
\ka\ lines, obtained using either keepcool=yes or =no. The comparison
shows the consistency within the statistical errors (see the next
section). In fact, the model fits with the keepcool=no spectra tend to
be slightly worse than the keepcool=yes spectra (as obtained
with the above described extra-cleaning). 

Table~\ref{t:xmm} also includes the orbital phases of individual observations.
For each observation, the phase is calculated for the middle time of the exposure duration 
with the ephemeris provided by Casares et al. (2009): an inferior conjunction reference 
time $T_{0}=2451387.148$ and an orbital period of $P=9.84450 \pm 0.00019$ days.
The error in the period can cause an uncertainty in the phase estimate up to $\sim 0.01$ (i.e., for
the latest observation), which is deemed insignificant for our analysis.

\section{{\it{XMM-NEWTON}} DATA ANALYSIS AND RESULTS}

\begin{figure}
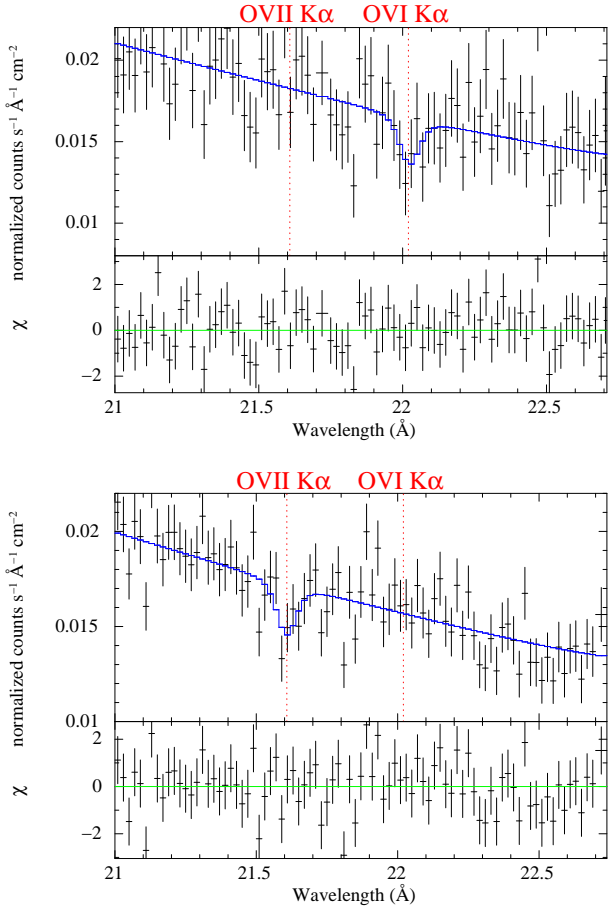

\includegraphics[height=3.3in,angle=270]{fig/f2a.ps} 
\includegraphics[height=3.3in,angle=270]{fig/f2b.ps} 
\caption{Missing \ovii\ K$\alpha$ line in N4 (top) and missing \ovi\ K$\alpha$ line in N6 (bottom).}
\label{f:spec_sample}
\end{figure}

The presence of the 
\ovii\ K$\alpha$ and \ovi\ K$\alpha$ lines in the combined spectrum is apparent (Fig.~\ref{f:spec_c};
last row in Table~\ref{t:xmm}). 
The shapes of these lines are consistent with the instrument LRF. Although 
the $\chi^2/\nu =201/165$ indicates that the fit is reasonable (which cannot be rejected 
at a confidence $\gtrsim 2\sigma$; Table~\ref{t:xmm}), the combined spectrum shows apparently 
residual systematic peaks and valleys. None
of these marginal deviations can be attributed 
to any known atomic transition, and 
 most of them are in the vicinities of 
the bad columns (as indicated by narrow dips of the model profile in 
the top panel of Fig.~\ref{f:spec_c}). In individual spectra, 
the features appear less obvious. 
The \ovii\ and \ovi\ K$\alpha$ absorption lines are also detected in the bulk of the individual observations at
various significance levels (Table~\ref{t:xmm}), but are clearly missing in the observations N4 (for
\ovii; upper panel in Fig.~\ref{f:spec_sample}) and N6 (\ovi; lower panel). In addition, the \ovi\ \ka\
line detection in N3 is also very marginal (at $\sim 1\sigma$). For
these three spectra, the EWs and their errors were 
measured with the centroid wavelengths of the respective lines
fixed to the corresponding best-fit values of the 
combined spectrum (Table~\ref{t:xmm}). 

\begin{figure}
\includegraphics[width=3.3in]{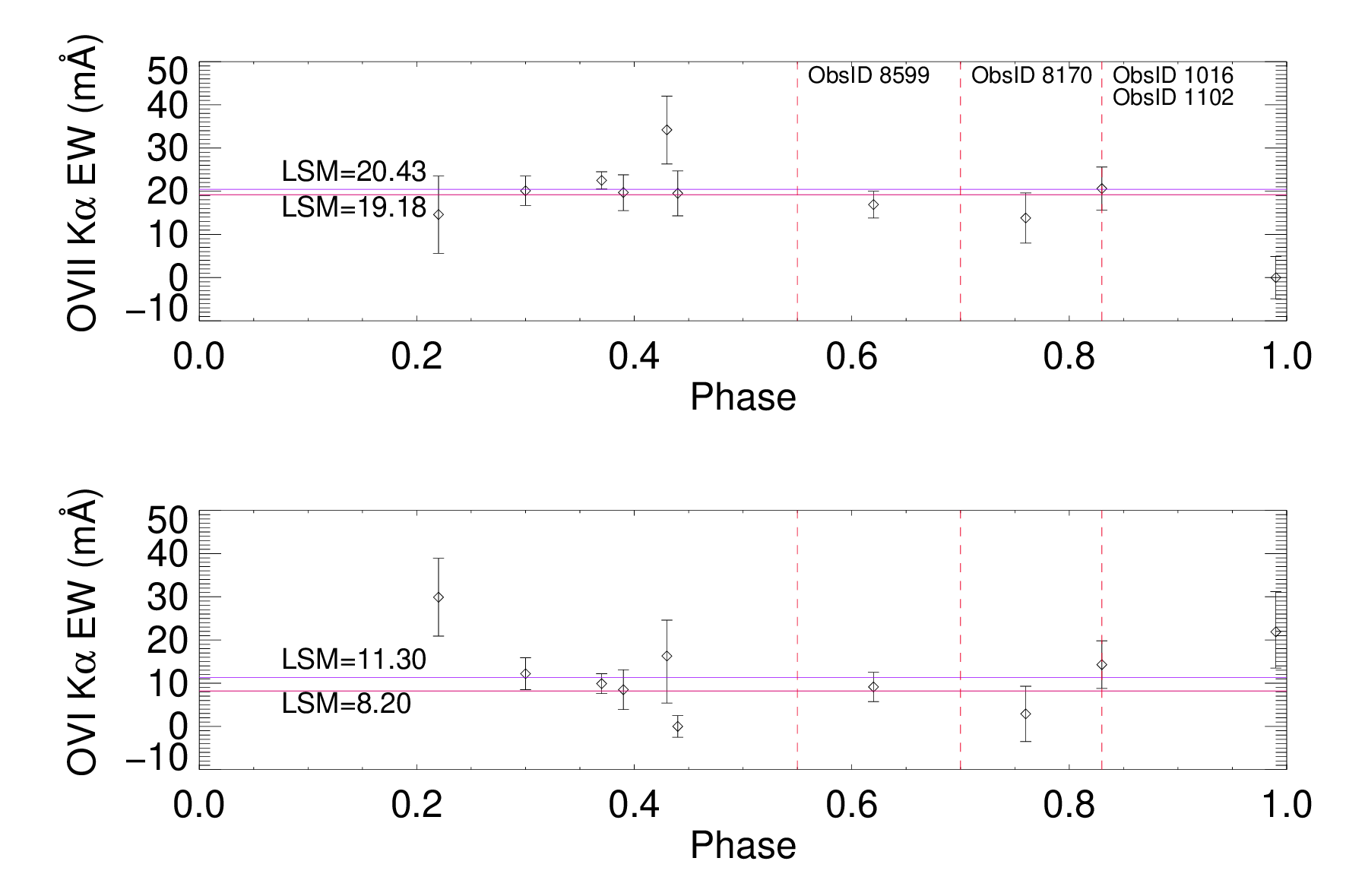} 
\caption{EWs of the \ovii\ and \ovi\ \ka\ lines vs. orbital phases for individual observations. 
Also plotted are the least-$\chi^2$ means with (upper blue) and without (lower purple) the non (or weak)-detections (data points with marginal or zero EWs). 
For reference, phases of {\it{Chandra}} exposures are marked by the vertical dashed lines.}
\label{f:EW_phase}
\end{figure}

We calculate the least $\chi^2$ mean of the EWs and test the null hypothesis of a constant EW for each line.
The least $\chi^2$ mean of the EWs (i.e., averaged with
weights $1/\sigma_i^2$, where $\sigma_i$ refer to the measurement errors from 
individual observations) is $19.18$~m\AA\ for \ovii\
and $8.20$~m\AA\ for \ovi\ for all the observations. These mean values are well
consistent with the EWs from the direct fit to the combined N1-10 spectrum.
With the mean values, the null hypothesis with $\chi^2/\nu =23.57/9$ and $23.39/9$ (for the \ovii\ and \ovi\ \ka\ lines)
can be ruled out at $\sim 99\%$ confidence for both lines. 
Excluding N3 (with no detection of \ovii) or N4 and N6 (no significant \ovi), 
the least $\chi^2$ mean of the EW becomes $20.43$~m\AA\ 
for \ovii\ \ka\ or $11.30$~m\AA\ for \ovi\ \ka, and the null hypothesis 
with $\chi^2/\nu =7.25/8$ and $7.48/7$ cannot be ruled out at any 
reasonable confidence. 
These different mean values, as well as the individual EWs, are shown 
in Fig.~\ref{f:EW_phase}. 
The EW of the \ovii\ or \ovi\ \ka\ lines in N4 or N6 deviates from the 
respective mean at the $\sim 4 \sigma$ 
or $3\sigma$ level. 
 N3,  N4  and N6 represent the shortest original exposures (Table~\ref{t:xmm}), but after cleaning N7 and N9 are shorter (Table~\ref{t:xmm}); see our  later discussion  of  potential exposure effects. 
 
\begin{figure}
\centering
\includegraphics[width=3.5in]{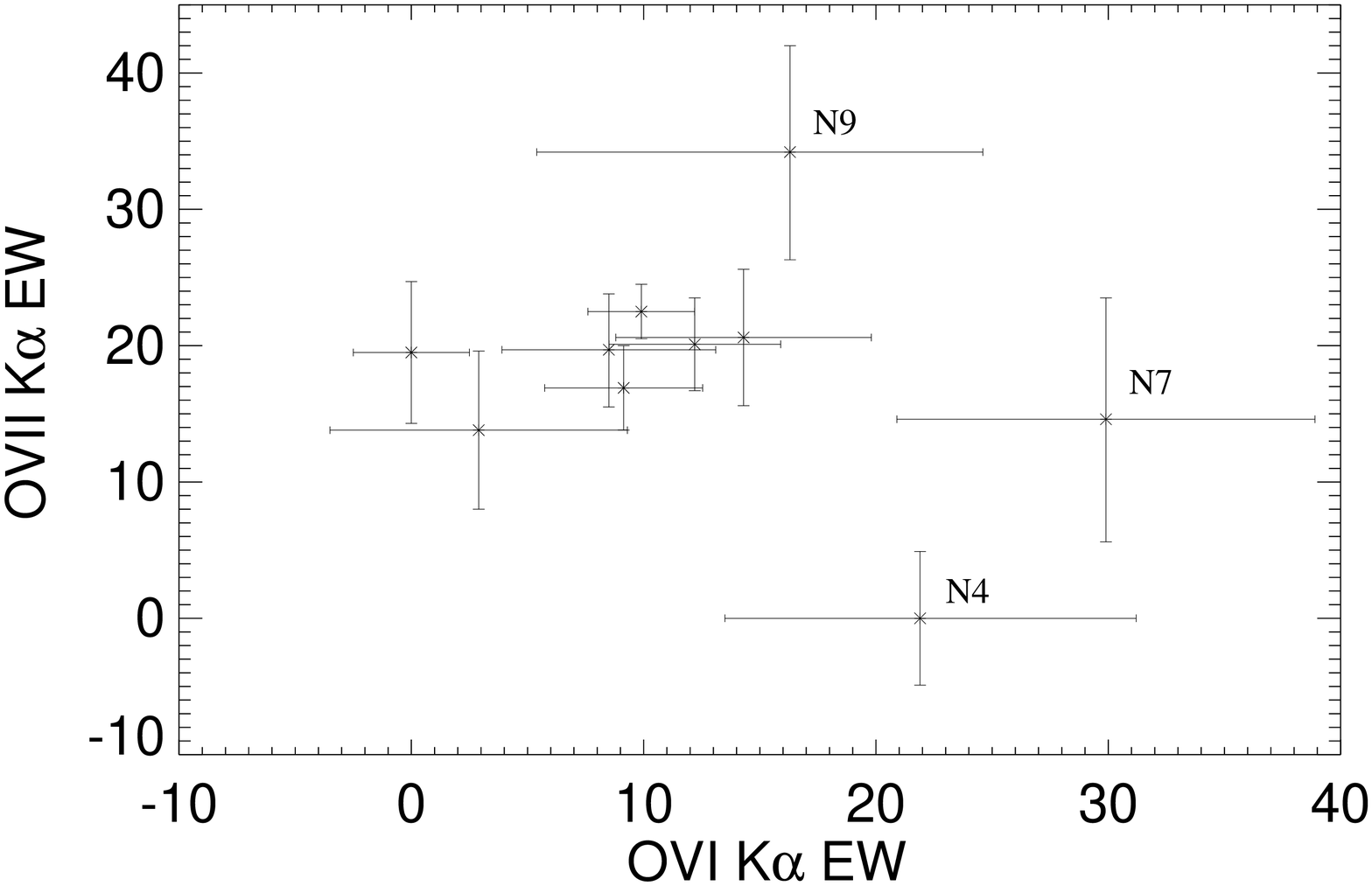}
\caption{OVII K$\alpha$ vs. \ovi\ K$\alpha$ EW plot. Three data points
from the shortest exposures are labeled.}
\label{f:EW-EW}
\end{figure}

A casual look of Fig.~\ref{f:EW_phase} may indicate a possible anti-correlation between the \ovii\ and \ovi\ \ka\ line EWs. At $\phi \approx 0.5$,
the EWs of the \ovii\ line tend to be larger, while those of the \ovi\ line tend 
to be smaller than at $\phi \approx$ 0 or 1.
This anti-correlation may be a bit more apparent in Fig.~\ref{f:EW-EW}, where the two EW measurements of the two lines are directly
compared. However, the scatter among the data points is large, in addition to the large error bars of individual measurements. With a bootstrapping re-sampling, accounting for both the scatter and the errors of individual measurements, the 
Spearman's Rank coefficient is $-0.018^{+0.37}_{-0.37}$, indicating that the anti-correlation 
is not significant.

\begin{figure}
\includegraphics[width=3.5in]{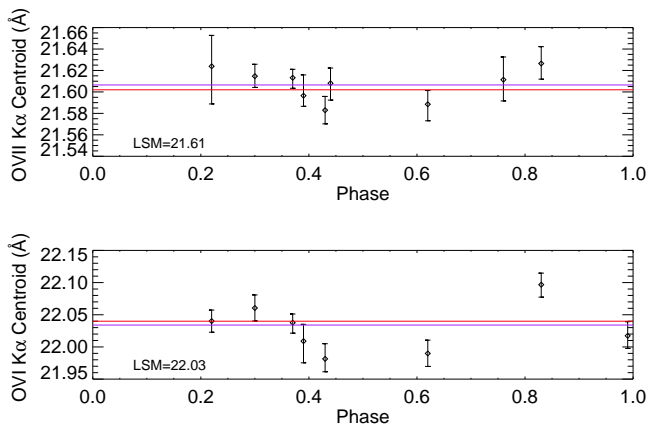} 
\caption{Absorption line centroid vs. orbital phase for the \ovii\ and \ovi\ \ka\ lines. 
Data points with fixed centroids in the fits are not included. The
blue horizontal lines represent the least-square means (as given in the
panels), while the red ones mark
the rest-frame wavelengths. 
} 
\label{f:w_phase}
\end{figure}

Similarly, we check the variability of the absorption centroids 
of the two absorption lines and the potential dependence on 
the orbital phase (Fig.~\ref{f:w_phase}). 
From the measured centroid wavelengths in Table~\ref{t:xmm},  we obtain their $\chi^2$ mean value as $21.607\pm0.005$ 
and $22.034\pm0.007$~\AA\ for the \ovii\ and \ovi\ \ka\ lines,
indicating no systemic velocity shifts from the rest-frame wavelengths ($69\pm69$ and $-81\pm95 {\rm~km~s^{-1}}$). 
The root mean square (RMS) deviation of the measurements are 0.012 m\AA\ and 0.020 m\AA\ for the two lines, 
corresponding to velocity dispersions
of 163 and 272 ${\rm~km~s^{-1}}$. It should be noted, however, the sampling of the orbit by the observations 
is probably too small to avoid a bias in the estimates of both the systematic velocities and the RMS deviations. 
Furthermore, while the centroids of the stronger \ovii\ line appear to be consistent with a constant 
(with the null hypothesis $\chi^2/\nu =8.72/8$), the \ovi\ line centroid shows significant variation 
($\chi^2/\nu =25.02/7$; at a confidence of 99.925\%).

\begin{figure}
\includegraphics[width=3.5in,angle=0]{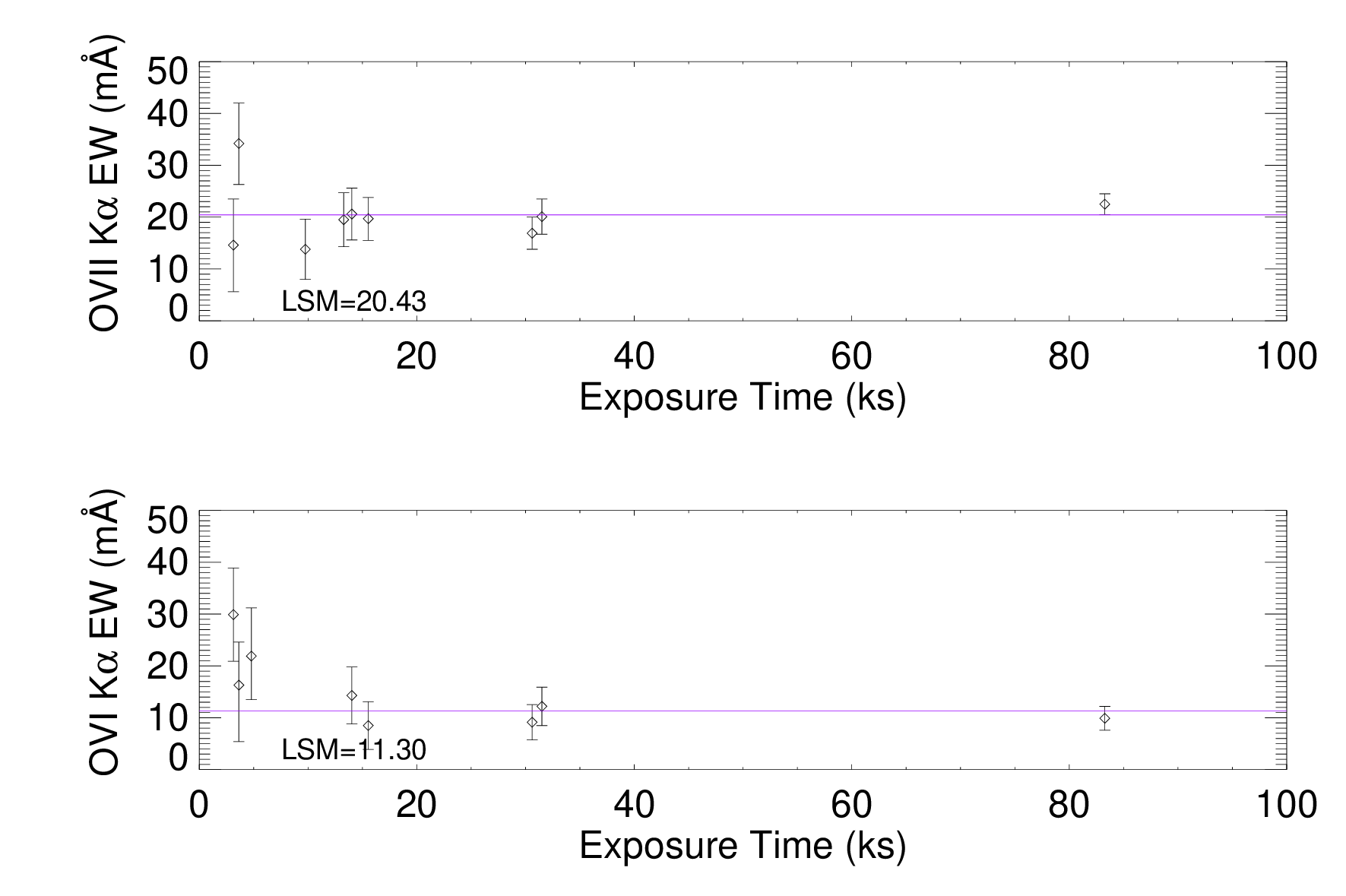} 
\includegraphics[width=3.5in,angle=0]{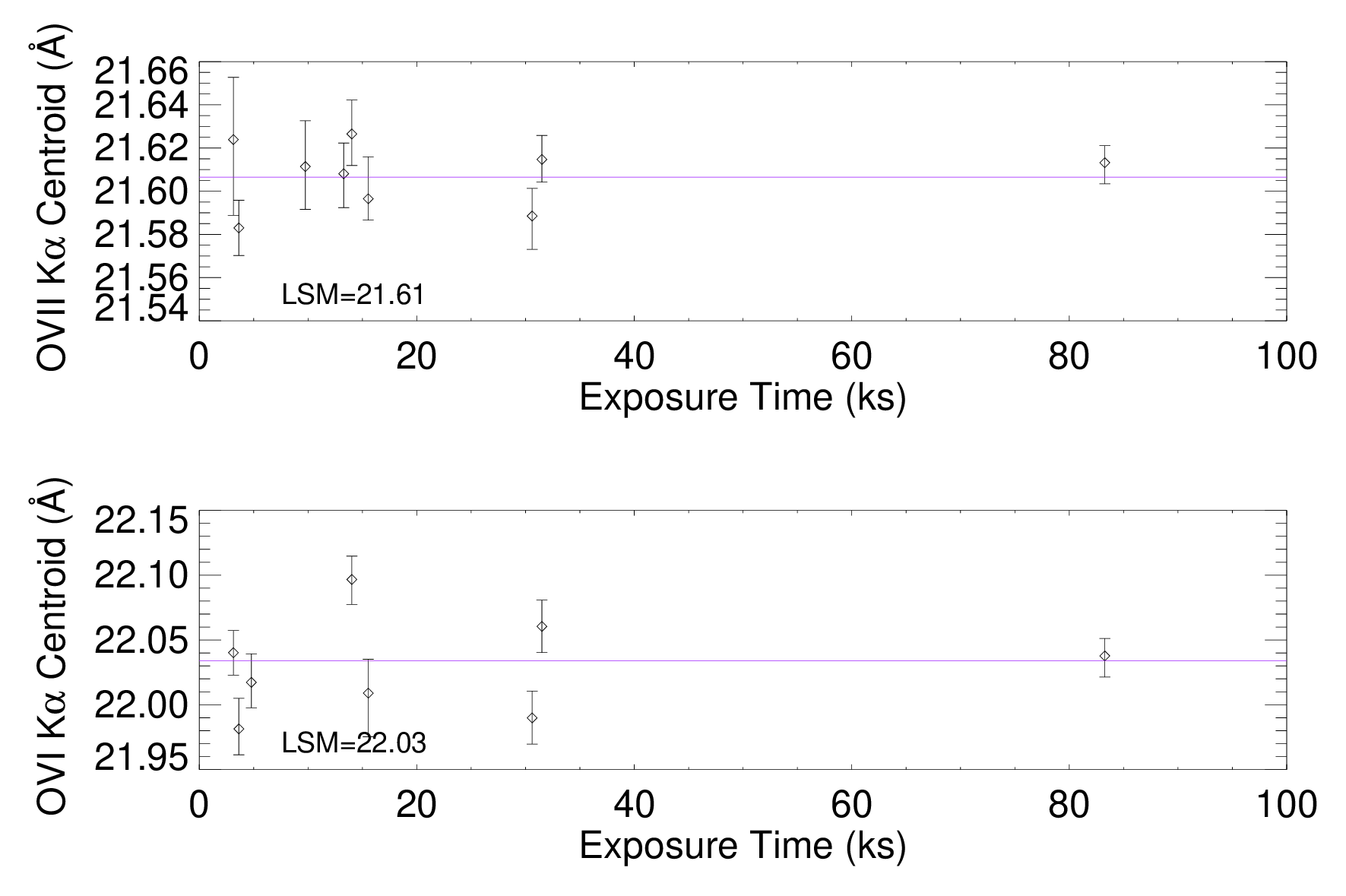} 
\caption{EWs and centroids of the \ovii\ and \ovi\ \ka\ lines vs. 
exposures of individual observations. 
Data points with zero EWs are not included. The
horizontal lines represent the least-square means. 
} 
\label{f:w_c_exp}
\end{figure}

One may be concerned that the measurements of the absorption lines may be affected
by the exposure of an observation. Indeed, the observation (N4) with zero
\ovii\ \ka\ EW has a relatively short 
exposure, although it is not the shortest (among the cleaned
exposures). But the
observation (N6) with zero \ovi\ \ka\ EW actually has a quite long cleaned
exposure (13 ks). Fig.~\ref{f:w_c_exp} shows no significant
dependence of the EWs or the centroids on the exposures. We label
the three data points obtained from the shortest
exposures ($< 5$ ks) in Fig.~\ref{f:EW-EW}. 
The remaining points, though with smaller error bars, still 
do not show any clear trend. 

\begin{figure}
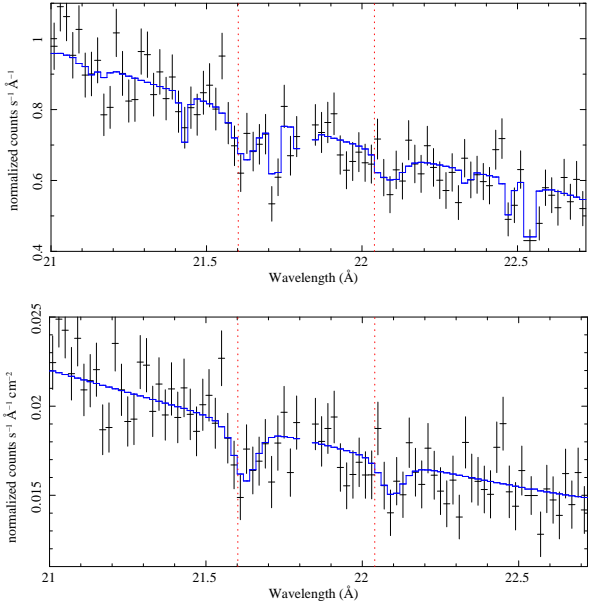

\centering
\includegraphics[width=1.5in,angle=270]{fig/f7a.ps}
\includegraphics[width=1.64in,angle=270]{fig/f7b.ps}
\caption{{\sl Upper panel:} Spectrum extracted from the \xmm\ grating observation N1 of Cyg X-2 together with the best-fit (blue curve) model. {\sl Lower panel:}
The same spectrum and fit, but approximately 
corrected for the effective area of the instrument. Various instrumental
features can be easily located in the spectrum of the upper panel, when compared with
that of the lower panel. }
\label{f:n1_inst}
\end{figure}

Much of the apparent centroid variation of the \ovi\ line (and
possibly the \ovii\ line as well) appears to be related to some bias
caused by the presence of instrumental features in the spectra. Most
of the spectra have such features in the immediate vicinities of the
fitting line. Although the features are supposedly
accounted for by the effective area correction, their presence could
still affect a fit (e.g., via an uneven counting statistics weighting), leading to
an apparent shift of the line centroid. This effect seems to be 
important for line fitting with poor counting statistics (e.g., to the \ovi\
\ka\ line) and is realized in various simulations with the XSPEC {\sl
  fakeit} routine. The effect is sometimes
apparent in the fits to the real spectra. Fig.~\ref{f:n1_inst}
illustrates an extreme case of such centroid shifts: the apparent
redshift of the 
\ovii\ line centroid (from the rest-frame wavelength) seems to be due to the
presence of an instrumental feature on the long wavelength side,
whereas the origin of the large redshift of the \ovi\ line centroid is
much less clear, although an instrumental feature is apparent at the
rest-frame wavelength of the line. Because of such uncertainties, the
\xmm\ centroid measurements for individual spectra are not used in the
rest of this paper. 

\section{COMPARISONS WITH OTHER OBSERVATIONS}

The \xmm\ detection of the \ovii\ absorption line and the
variation of both \ovii\ and \ovi\ line EWs prompt us to re-examine the
\chandra\ observations of Cyg X-2 and to extract relevant information
on both line and
continuum emissions of the source from observations
made with \hst\ and \rxte. The results are then compared with those obtained from \xmm\ observations.

\subsection{Re-examination of the {\it Chandra} data}

\begin{figure}
\centering
\includegraphics[width=3.2in,angle=0]{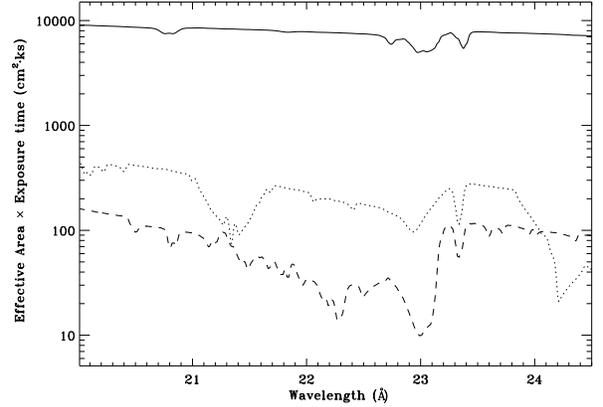}
\caption{Effective Area of {\it Chandra} HETG observations MEG:
the negative order (dotted line) and the positive order (dashed).
The RGS1 effective area (solid) is included for comparison.
}
\label{f:ea_chandra}
\end{figure}

We re-examine the validity of the existing \chandra\ results,
based on a careful checking of various potential instrumental effects 
in our concerned wavelength range. We check whether or not the absence of the \ovii\ 
K$\alpha$ line in the \chandra\ data could be due to instrumental artifacts and 
quantify the inconsistency of the data with our \xmm\ measurement.
As shown in Fig.~\ref{f:ea_chandra}, the effective areas 
of the RGS1 is about a factor of 40 greater than that of the MEG at the 
\ovii\ \ka\ wavelength  (Fig.~\ref{f:ea_chandra}). But the spectral resolution of the MEG is significantly 
better than that of the RGS1. In terms of the figure of merit for detecting
the \ovii\ \ka\ line, the RSG1 is still
about a factor of 5 more effective than the MEG.

Following Yao et al. (2009), our re-examination is based primarily
on a co-added spectrum of four High Energy Transmission Grating (HETG) observations
(Table~\ref{t:chandra}). 
We use only the Medium Energy Grating parts for their relatively
high sensitivity.  The four observations were taken
with a SIM-Y offset of 1.167 arc-minutes. 
With such a configuration, the gap between the front-illuminated (FI) CCDs
(S4 and S5) is at $\sim 22$~\AA\ in 
positive order spectrum\footnote{http://cxc.harvard.edu/cgi-gen/LETG/alp.cgi},
while the
back-illuminated (BI) CCD (S1) covers the wavelength range of 
12.5-24.0~\AA\ in the negative order MEG spectrum (Fig.~\ref{f:ea_chandra}).
At $\sim24$~\AA, the significant drop of the effective area of the negative
order MEG is consistent with this
configuration. However there is also a remarkable drop around 21.4~\AA\ with a
width of 0.8~\AA, which could be due to numerous bad columns and pixels that have
been excised in the continuous-clocking  observation mode along with
the dithering effect. Unfortunately, our line of interest, \ovii\ K$\alpha$, 
landed in this region. Nevertheless, at wavelengths around 21.6~\AA, 
the BI CCD has more than twice the effective
area than the FI CCD, and dominates the contribution to the spectral
counts. Therefore, we examine the \ovii\ K$\alpha$ absorption line in
the spectra extracted from the negative order alone. 

\begin{figure}
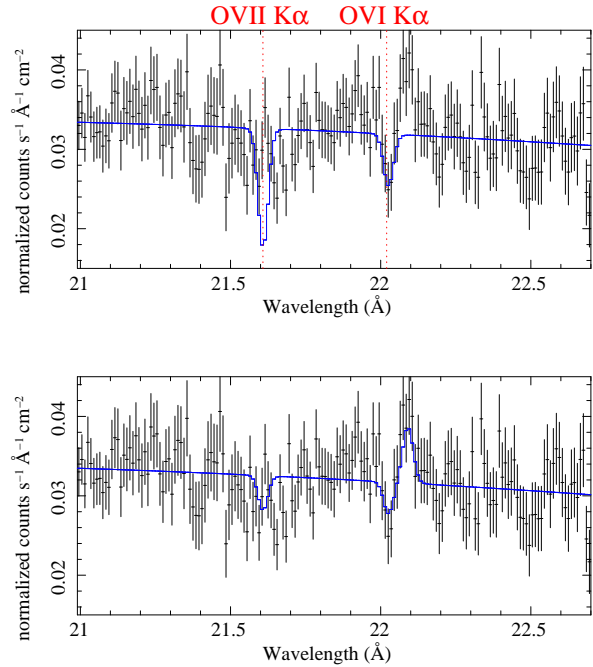

\includegraphics[width=1.8in,angle=270]{fig/f9a.ps}
\includegraphics[width=1.8in,angle=270]{fig/f9b.ps}
\caption{{\sl Upper panel:} 
The combined MEG spectrum with the adopted model for the simulations 
described in the text. The model consists of the best-fit continuum to the 
spectrum (see the lower panel), together with the \ovii\ and \ovi\ 
\ka\ absorption lines from the best-fit to the combined RGS spectrum.
{\sl Lower panel:} The same spectrum with the  best-fit model, consisting of
the power law continuum plus the two absorption lines, as well as an 
emission line.}
\label{f:Model_Meg}
\end{figure}

To check the consistency of the mean \ovii\ \ka\ absorption line
measurements, we simulate 1000 MEG spectra using the best-fit model
of the line in the RGS combined spectrum, while the continuum is taken as
measured in the real co-added MEG spectrum, which gives the 95\% upper limit of 
the EW as 6.3 m\AA\ (Yao et al. 2009). Fig.~\ref{f:Model_Meg} upper
panel shows an overlay of the 
adopted model on the spectrum. Each simulated spectrum is fitted to obtain
the line strength. We find that the smallest strength value is about
10 m\AA. Therefore, the consistency of the MEG upper limit with the
RGS measurement can be ruled out statistically at a greater than 99.9\% confidence. 

We further conclude that this non-detection of the line in the MEG spectrum
cannot be attributed to any known
instrumental effects. The shape of the effective area used seems to be quite 
reasonable; the spectra near the line can be well fitted with a simple 
power-law model, as would be expected (Fig.~4 in Yao et al. 2009). We also find no clear
dependence of such missing absorption lines on the orbital phase of
the source, based on analysis of the MEG spectra from individual observations. As shown in Fig.~\ref{f:EW_phase}, one of the \chandra\
observations coincides in phase (though not observing date) with an
\xmm\ observation which has a clear detection of the \ovii\ line.
It thus remains puzzling as to why the absorption line was absent in all 
of the four \chandra\ observations taken on different dates and/or in 
different orbital phases (Table~\ref{t:chandra}), while the line
disappears only occasionally in \xmm\ observations. Admittedly,
we cannot completely rule out the consistency when a pair 
of observations are compared with each, especially when the counting
statistics is poor (e.g., due to relatively short exposure) and hence
error bars are large. Such a comparison will thus not be particularly useful.
What we can firmly conclude is that the absorption line strength is not 
a constant.

The \ovi\ \ka\ absorption line as detected in the MEG spectrum has an EW of 10.8 ($\pm 2.4$ m\AA) (Table 4 of Yao et al.), statistically consistent with the
mean EW of 11.3 m\AA\ obtained with RGS spectra, excluding the non-detection 
observation N6. Interestingly, the close-up of the spectral range as shown in 
Fig.~\ref{f:Model_Meg} indicates the presence of an emission line at 
$22.086^{+0.007}_{-0.012}$ \AA, just on the
longer wavelength side of the \ovi\ absorption line. To assess its significance, we
fit the line with an additive Gaussian (Fig.~\ref{f:Model_Meg}). The fitted dispersion and EW are
$3.4_{-1.5}^{+2.1} \times 10^{-4}$ keV and $9.6_{-3.4}^{+2.9}$ m\AA,
which are all comparable to that of the  absorption line. Therefore, this emission line may be responsible for
the occasional disappearance of the absorption line, when the two are blurred together.

We also check the consistency of other absorption lines detected by Yao et al.
(2009). Unfortunately, the spectral regions around all these lines are in general
severely affected by various instrumental features
in the RGS spectra. We find it difficult to obtain conclusive results. 
In particular, the \ovii\ K$\beta$ line centroid 
measured in the MEG spectrum is significantly blue-shifted 
relative to the rest-frame wavelength (18.654 \AA) calculated by Ming F. Gu, 
but is consistent with the value (18.628 \AA) given in the catalogs of 
NIST and Verner et al. (Table 1 of Yao et al.). 
A similar offset (though smaller) is also present for \neix\ K$\beta$ line. 
It seems that Gu's calculation may be systematically biased for the K$\beta$ 
transitions of He-like ions. The detection of these lines appears to be rather secure 
 in the  MEG spectrum (with high detection significance and 
no obvious instrumental effect). With the caveats of multiple instrumental
features (which may not be adequately corrected) 
in the spectral region of the RGS data and assuming the 
wavelength of 18.628 \AA, we obtain a measurement of
the line with an EW  of about $4.7 \pm 1.5$ m\AA~(1$\sigma$ errors), which
is consistent with the MEG measurement of $4.4\pm0.8$ (Table 4 in Yao et al. 
2009).

\begin{table}
\tiny
\centering
\caption{{\it Chandra} observations of Cyg X-2}
\label{t:chandra}
\begin{tabular}{@{}lccccr@{}}
\hline\hline
Obs\# & ObsID & Exposure (ks) & Start Time & $\phi$ & $f_{ASM}$\\ [0.5ex]
\hline
C1 &1102 & 29 &1999-09-23@21:12 & 0.83 & 20.78 \\
C2 &1016 & 15 &2001-08-12@03:42 & 0.83 & 45.87 \\
C3 &8599 & 71 &2007-08-23@05:01 & 0.55 & 29.07 \\
C4 &8170 & 79 &2007-08-23@17:44 & 0.70 & 44.94 \\
\hline
\end{tabular}
\end{table}

\begin{table}
\tiny
\centering
\caption{{\it HST}/COS observations of Cyg X-2}
\label{t:cos}
\begin{tabular}{@{}lccccr@{}}
\hline\hline
Obs\# & Exp. (s) & Start Time & $\phi$ & Flux\\ [0.5ex]
\hline
H1  &  7044.8   &         2010-09-21 14:11 &  0.81 &3.3\\
H2  &  7111.7   &         2010-09-24 10:53 &  0.11 &3.3\\
H3  &  7114.7   &         2010-09-26 09:13 &  0.32 &3.3\\
\hline
\end{tabular}

Note: The flux is in units of $10^{-15} {\rm~ergs~cm^{-2}~s^{-1}~\AA^{-1}}$.
\end{table}

\subsection{Emission lines as observed by {\it HST}/COS}

{\it HST} visited Cyg X-2 three times with the COS (Table~\ref{t:cos}). These observations were taken with the FP\_POS 
position shifting from 1 to 4 during each visit. We calibrated the observations
with the pipeline CALCOS (version 2.11f). Flat-fielding and alignment of the
processed exposures were carried out using IDL routines developed by the COS
science team\footnote{See http://casa.colorado.edu/$\sim$danforth/costools.html for the co-adding
and flat-fielding algorithm and additional discussion.}.
The extracted spectra from individual exposures of each observation
were cross-correlated and 
combined to form an exposure-weighted co-added spectrum.

\begin{figure*}
\includegraphics[width=6in]{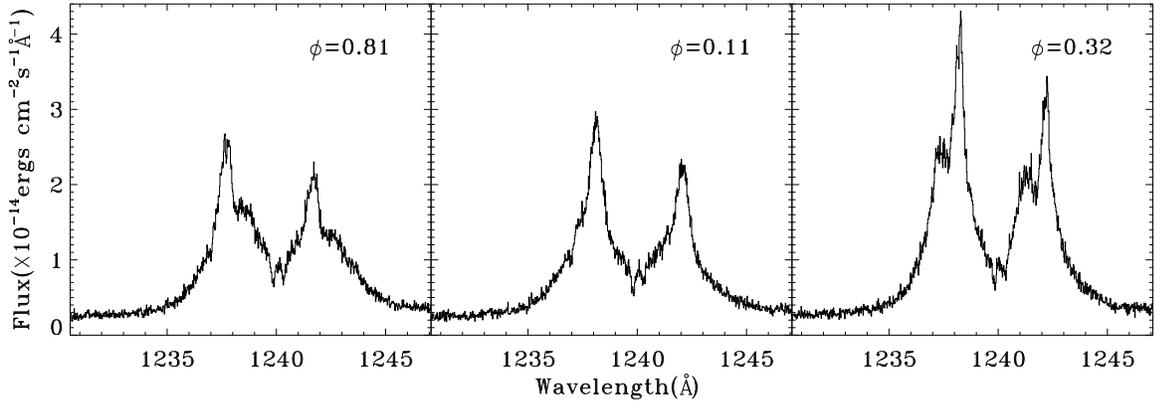} 
\caption{{\it HST}/COS spectra of the \nv\ doublet at three phases (marked in the panels).}
\label{f:cos}
\end{figure*}

Here we focus on the \nv\ doublet region of the COS spectra (Fig.~\ref{f:cos}). 
Compared to the rest-frame wavelengths of 1238.82 \AA\ and 1242.81 \AA,
the line centroids are blue-shifted, consistent with similar
velocities 
as seen in the GHRS spectra (Vrtilek \etal\ 2003).
The profile of each line apparently consists of two major components:
one (with higher peak fluxes) seems to be considerably narrower than
the other. While a quantitative decomposition and modeling of these
components are beyond the scope of the present work, a few trends are
readily noticeable. Both the centroid and flux of these two components
change systematically with the phase. The positive centroid velocity
shift of the narrow component, as seen from the left to the right
panels, is qualitatively consistent with its association with the companion, as tentatively suggested by Vrtilek et al. (2003). 
The component appears substantially narrower than inferred from the GHRS spectrum (FWHMs $\approx 316$ and $336 {\rm~km~s^{-1}}$) and is again more
consistent with the companion origin than with the accretion disk. 
The velocity shift of the broad component with the phase appears to be in the
opposite direction of the narrow one. Therefore, the broad component can be attributed naturally to the accretion disk. Both components appear to be the
brightest in the $\phi = 0.32$ plot. This trend for the broad component may be consistent with the saddle shaped disk model, in which less obscuration
is expected at $\phi \approx 0.25$ or 0.75. For the narrow component, some obscuration may be expected from the companion itself, particularly 
at $\phi \approx 0$. With an inclination of 62.5$^\circ$ for the binary system
(Casares et al. 2009), the line emission may originate in an outflow from the heated surface of the companion.

\subsection{Correlation with broad-band X-ray fluxes}

\begin{figure}
\includegraphics[width=3.3in]{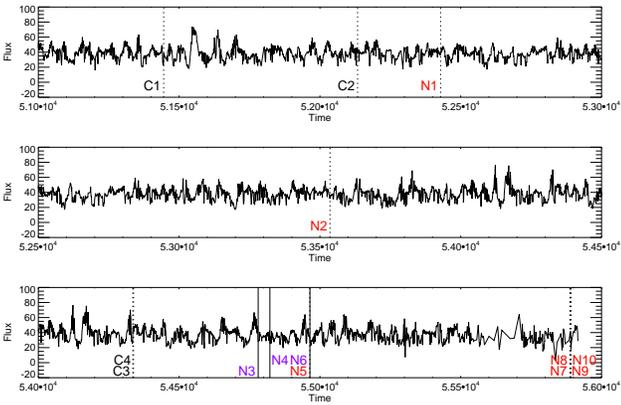}
\caption{Light-curve of Cyg X-2: the \rxte\ /ASM flux in the 1.5-12 keV band. The 
times for the {\it{XMM-Newton}} and {\it{Chandra}} 
exposures are marked by the vertical lines in red and black. Blue \xmm\ exposures (N3, N4 and N6) had very weak or no detection of either the 
\ovii\ and \ovi\ K$\alpha$ line.}
\label{f:w_rxteflux}
\end{figure}

We further explore the potential connection of the line absorption with the broad-band
X-ray fluxes of Cyg X-2. In Fig.~\ref{f:w_rxteflux}, for example, we
have marked all the grating observations on the light curve obtained
with the \rxte\ /ASM. 
However, we find no apparent correlation of the absorption line EWs seen in the RGS or HETG 
observations with the ASM or RGS broad-band fluxes  (Tables~\ref{t:xmm} and~\ref{t:chandra}).

\section{DISCUSSION}
What is the origin of the \ovii\ and \ovi\ \ka\ absorption lines? 
Along the line of sight toward Cyg X-2 (Galactic coordinates $l =87.3^\circ, b=-11.3^\circ$), a typical ISM absorber
is expected to have a systemic velocity in the range of $\sim 0$ to $-100~{\rm km~s^{-1}}$, as seen in 
21 cm neutral hydrogen surveys. In contrast,
the systematic velocity of Cyg X-2 is known to be $-209.9\pm1.4
{\rm~km~s^{-1}}$ (Casares \etal\ 2009), which means that its emission
at the systemic velocity would be $\sim $15 m\AA\ off from the absorption line
centroid; the line-of-sight orbital velocity of the companion,
$\lesssim 87{\rm~km~s^{-1}}$, alone can 
hardly make a difference. 
The measured 
centroid of the \ovii\ \ka\ line, $21.608\pm0.004$ \AA, appears to be
slightly redshifted, compared to the expected rest-frame wavelength of
21.602 \AA.
But the measured line centroid of the
\ovi\ \ka\ line, $22.026\pm0.009$ \AA, compared to the expected 22.040
\AA, shows an opposite shift. One also needs to consider the systematic 
absolute wavelength uncertainty of 7~m\AA\ (1$\sigma$), as is expected for an 
RGS observation. 
We add the statistical and systematic uncertainties in quadrature and find that the centroids 
of both lines are consistent with an ISM origin (within
$\sim 2\sigma$) and that the \ovii\ line is significantly redshifted from the
systemic velocity of Cyg X-2 at a confidence level of $\sim 2.6\sigma$. 

If the absorption lines arise from the hot ISM, then the observed
variation in the EWs must be related to the changing intensity of the
emission lines associated with Cyg X-2.  
Because of the limited spectral resolution of 
the RGS (FWHM $\approx$ 60 m\AA), 
the emission and absorption lines should be blended. But a velocity offset
between the line emission and absorption could also be large enough to
affect the centroid measurement of the absorption line. However, as
mentioned in \S~3, the
current measurements are not sufficiently reliable to constrain the
emission properties of the source.

Another interesting issue that needs an explanation is
the fact that the \ovii\ K$\beta$ absorption line was
detected, while the \ovii\ \ka\ line was missing in the same \chandra/HETG spectrum
(Yao \etal\ 2009). This may be understood in our emission filling scenario because the emissivity of
the K$\alpha$ transition is substantially higher than that of the K$\beta$
transition for a plasma of $\sim 10^6$ K. In addition, the
effectiveness of the emission filling can be affected by the substantial
difference in the absorption saturations of the two lines. The
optical depth is $\sim 2.5$ at the center of the K$\alpha$ line and
$\sim 0.5$ for the K$\beta$ line for an absorbing plasma of a column
density, log[N(\ovii)] $= 16$, and a velocity dispersion, $b=100
{\rm~km~s^{-1}}$ (Yao \etal\ 2009). In contrast, there is no saturation for the
corresponding emission lines. Thus it is easier to fill a highly
saturated \ovii\ \ka\ line than a weakly saturated \ovii\ K$\beta$ line.

What is the origin of the emission lines?
Possible sites are the accretion disk, the impact bulge of the
accretion stream, and the surface or even outflow of the stellar
companion. Emission lines from the first two sites should generally be broad
(as observed in those short wavelength lines; Schulz \etal\ 2009), due 
primarily to the fast rotation around the neutron star, depending on the exact location of the emission site.
In comparison, the emission from the companion should be relatively
narrow. Therefore, the \ovii\
and \ovi\ \ka\ lines, as well as the
narrow component of the \nv\ doublet seen in the COS spectra, are
likely to be associated with the companion or its wind.
Also the relative strengths of these emission lines are sensitive to the 
temperature of X-ray-emitting plasma. Therefore, the lines may not
arise from the same locations. So a correlation between the line emissions
is not necessary, which may explain the lack of the coincidence of the missing \ovii\ and
\ovi\ absorption lines (\S~3).  Indeed, the redshift of the emission line centroid wavelength
corresponds to a velocity shift of $\sim 6 \times 10^2
{\rm~km~s^{-1}}$, relative to the rest-frame
wavelength (22.0403 \AA) and larger to the blue-shifted Cyg X-2. This large velocity shift, if real, indicates an association of the
emission line with the accretion flow. However, this scenario can
hardly explain the narrowness of the line. The emission and
absorption lines together resembles a P-Cygni profile, which could
originate in a thick wind. However, the Cyg X-2 has a low-mass
companion, which is not expected to have a strong wind. Also for a wind-generated
P-Cygni profile, the absorption is expected to be strong
blue-shifted, while the emission should peak at the rest-frame
wavelength, which is opposite to what is observed for Cyg X-2. So
the origin of the emission line remains largely unknown.

The \chandra\ or \xmm\ grating observations were taken in various
different orbital cycles, in which Cyg X-2 may be in very different
states. Therefore, the observations probably sampled quite different accretion, heating, and radiation processes; hence different X-ray line emissions, even if the
binary had similar orbital phases and/or continuum X-ray fluxes (Ba\l
uci\'nska-Church \etal\ 2011; Vrtilek \etal\ 2003).
Therefore, it is highly desirable to have a set of dedicated X-ray
spectroscopic observations in a single orbital cycle. As
illustrated in \S~3, potential instrumental effects on the measurements of
the absorption lines should also not be neglected. In addition to 
the discussed effect on the centroid measurement, the 
presence of instrumental features, which may not be corrected
accurately, could, in principle, affect the quantitative
measurements of the EW of a line as well. The pointing
direction of the observations should thus be set carefully to minimize the presence of the
instrumental features in the immediate vicinities of the key
absorption lines.

\section{SUMMARY}

Motivated by solving the mystery of the missing \ovii\ \ka\ absorption in a previous \chandra/HETG study, 
we have systematically analyzed 10 \xmm/RGS observations of Cyg X-2 in the wavelength range covering both 
the \ovii\ and \ovi\ \ka\ lines. This analysis is complemented by both a 
re-examination of the \chandra/HETG data 
and an analysis of 
the \nv\ 1238.82~\AA\ and 1242.81~\AA\ doublet in three \hst/COS spectra.
The main results and conclusions that we have obtained are as follows:

\begin{itemize}

\item We have clearly detected the \ovii\ \ka\ absorption line in the bulk of the RGS observations. 
The mean EW of the line is $19.6 \pm 1.3$ m\AA. But in one of the
observations, the lack of the line (or an inconsistency
with the mean EW at the $\sim 4\sigma$ level) is also detected, similar to the absence of the line in the \chandra/HETG spectra, which is confirmed in our re-analysis. 
Similar EW variation is also detected for the \ovi\ \ka\ line. But this variation is not correlated with that of the \ovii\ \ka\ line. 

\item The mean centroids of the two absorption lines are consistent
  with their origin in the ISM. The EW variations in the lines are
  thus likely caused by changing contamination of the corresponding
  narrow line emission within the LRF of the RGS. 

\item A re-examination of the MEG spectrum shows an apparent narrow
emission line right on the longer wavelength edge of the \ovi\ \ka\
absorption line. Their EWs are also comparable. Therefore, such an
emission, when observed with a lower spectral resolution and/or shifted to a
shorter wavelength, could be responsible for the occasional disappearance of the
\ovi\ \ka\ absorption in the RGS spectra. 

\item A narrow component of the \nv\ emission doublet is also seen in
  the \hst/COS spectra of Cyg X-2. This component, likely associated with the X-ray heated
binary companion, varies in the line strengths and centroids.

\item The exact origin of the putative \ovii\ and \ovi\ \ka\ line emission is yet
 to be understood.
A dedicated monitoring study of the X-ray line absorption over a
single orbital period could
be particularly useful to constraining the site of the line emission and hence 
understanding its nature. 

\end{itemize}

\section*{Acknowledgments}
We are grateful to the referee for valuable comments that lead to
improvements of the paper.
This research made use of \xmm\ archival data.
\xmm\ is an ESA science mission with instruments
and contributions directly funded by ESA Member States and the USA (NASA).
We thank Dr. Herman Marshall for comments on various HETG
calibration issues and gratefully acknowledge the support by NASA via
the summer research fellowships from the Massachusetts Space Grant
Consortium and the grants NNX10AE85G and AR1-12018X. Y.Y. acknowledges the funding support from \chandra\ archival program AR1-12002A.

\end{document}